\documentclass[aps,prl,twocolumn,amsmath,amssymb,footinbib,showpacs,longbibliography]{revtex4-1}

\newcommand{\Jnatphys}{Nat. Phys.}

\newcommand{\Jprl}{Phys. Rev. Lett.}
\newcommand{\Jpr}{Phys. Rev.}
\newcommand{\Jpra}{Phys. Rev. A}
\newcommand{\Jprb}{Phys. Rev. B}

\newcommand{\Jrmp}{Rev. Mod. Phys.}

\newcommand{\JRepProgPhys}{Rep. Prog. Phys.}

\newcommand{\Jphystoday}{Phys. Today}

\newcommand{\Jadvphys}{Adv. Phys.}

\usepackage[english]{babel}
\usepackage{latexsym}
\usepackage{graphics}
\usepackage{subfigure}
\usepackage{epsfig}
\usepackage{color}
\usepackage{hyperref}
\usepackage{braket} 
\usepackage[T1]{fontenc}
\usepackage[latin9]{inputenc}
\usepackage{amstext}
\usepackage{amssymb}
\usepackage{graphicx}
\usepackage{esint}
\usepackage{braket}
\usepackage{babel}
\usepackage{amsmath}

\hypersetup{
colorlinks=true,
citecolor=blue,
linkcolor=red,
urlcolor=black
}


\newcommand{\lettersection}[1]{\section{#1}}
\renewcommand{\lettersection}[1]{\paragraph*{#1.---}}

\newcommand{\ie}{i.e.}
\newcommand{\eg}{e.g.}

\newcommand{\IPR}{\textrm{IPR}}
\newcommand{\Er}{E_{\textrm{r}}}
\newcommand{\Ec}{E_{\textrm{c}}}
\newcommand{\Vc}{V_{\textrm{c}}}

\renewcommand{\DH}{D_\textrm{\tiny H}}
\newcommand{\NB}{N_\textrm{\tiny B}}

\begin{document}

\newcommand{\letitre}{Critical Behavior and Fractality in Shallow One-Dimensional Quasiperiodic Potentials}

\title{
\letitre
}

\author{Hepeng Yao}
\affiliation{
CPHT, CNRS, Ecole Polytechnique, Institut Polytechnique de Paris, Route de Saclay, F-91128 Palaiseau, France
}

\author{Alice Khoudli}
\affiliation{
CPHT, CNRS, Ecole Polytechnique, Institut Polytechnique de Paris, Route de Saclay, F-91128 Palaiseau, France
}

\author{L\'ea Bresque}
\affiliation{
CPHT, CNRS, Ecole Polytechnique, Institut Polytechnique de Paris, Route de Saclay, F-91128 Palaiseau, France
}

\author{Laurent Sanchez-Palencia}
\affiliation{
CPHT, CNRS, Ecole Polytechnique, Institut Polytechnique de Paris, Route de Saclay, F-91128 Palaiseau, France
}

\date{\today}

\begin{abstract}
Quasiperiodic systems offer an appealing intermediate between long-range ordered and genuine disordered systems, with unusual critical properties.
One-dimensional models that break the so-called self-dual symmetry usually display a mobility edge, similarly as truly disordered systems in dimension strictly higher than two.
Here, we determine the critical localization properties of single particles in shallow, one-dimensional, quasiperiodic models and relate them to the fractal character of the energy spectrum.
On the one hand, we determine the mobility edge and show that it separates the localized and extended phases, with no intermediate phase.
On the other hand, we determine the critical potential amplitude and find the universal critical exponent $\nu \simeq 1/3$.
We also study the spectral Hausdorff dimension and show that it is nonuniversal but always smaller than unity, hence showing that the spectrum is nowhere dense.
Finally, applications to ongoing studies of Anderson localization, Bose-glass physics, and many-body localization in ultracold atoms are discussed.
\end{abstract}

\maketitle

In an homogeneous system, all the single-particle wave functions are extended.
In contrast, they may be exponentially localized in the presence of disorder owing to the breaking of translational invariance~\cite{anderson1958}.
This effect, known as Anderson localization, is a fundamental, ubiquitous phenomenon at the origin of metal-insulator transitions in many systems~\cite{abrahams2010}.
Quasiperiodic models hold a special place for they are at the interface of long-range ordered and fully disordered systems.
They describe a variety of systems, including
quasicrystals~\cite{shechtman1984},
electronic materials in orthogonal magnetic fields~\cite{peierls1933,harper1955,hofstadter1976}
or with incommensurate charge-density waves~\cite{wilson1975},
Fibonacci heterostructures~\cite{merlin1985},
photonic crystals~\cite{lahini2009},
and cavity polaritons~\cite{tanese2014}.
They also proved pivotal in quantum gases~\cite{aspect2009,modugno2010,lsp2010} to investigate
Anderson localization of matter waves~\cite{damski2003,roth2003,roati2002}
and interacting Bose gases~\cite{lellouch2014},
the emergence of long-range quasiperiodic order~\cite{lsp2005,mace2016,viebahn2019},
Bose-glass physics~\cite{damski2003,roth2003,fallani2007,gadway2011,tanzi2013,derrico2014},
and many-body localization~\cite{iyer2013,lukin2018,matthew2018,kohlert2019}.

Anderson localization in quasiperiodic systems, however, significantly differs from its counterpart in truly disordered systems.
While in a disordered system a phase transition between the Anderson-localized and extended phases occurs only in dimension strictly higher than $2$~\cite{abrahams1979}, it may occur in one-dimensional (1D) quasiperiodic systems.
The most celebrated example is the Aubry-Andr\'e (AA) Hamiltonian, obtained from the tight-binding model generated by a strong lattice, by adding a second, weak, incommensurate lattice.
In the AA model, the localization transition occurs at a critical value of the quasiperiodic potential, irrespective of the particle energy~\cite{aubry1980}. This behaviur results from a special symmetry, known as self-duality.
When the latter is broken, an energy mobility edge (ME), \ie\ a critical energy separating localized and extended states, generally appears,
as demonstrated in a variety of models~\cite{soukoulis1982,dassarma1986,sarma1990,biddle2009,biddle2010,biddle2011,ganeshan2015,szabo2018}.
One of the simplest examples is obtained by using two incommensurate lattices of comparable amplitudes.
This model attracts significant attention in ultracold-atom systems~\cite{boers2007,li2017}.
They have been used to study many-body localization in a 1D system exhibiting a single-particle ME~\cite{kohlert2019} and may serve to overcome finite-temperature issues in the observation of the still elusive Bose-glass phase~\cite{derrico2014,gori2016} (see below).
Recently, the localization properties and the ME of the single-particle problem have been studied both theoretically~\cite{li2017} and experimentally~\cite{luschen2018}. However, important critical properties of this model are still unknown.
For instance, whether an intermediate phase appears in between the localized and extended phases remains unclear.

In this work, we study the critical properties and the fractality of noninteracting particles in shallow quasiperiodic potentials.
We consider various models, including bichromatic and trichromatic lattices with balanced or imbalanced amplitudes.
In all cases, above a certain critical amplitude of the quasi-periodic potential $\Vc$, we find a finite energy ME.
It marks a sharp transition between the localized and extended phases with no intermediate phase.
The ME is always found in one of the energy gaps, which are dense.
We show that this is a direct consequence of the fractal character of the energy spectrum, which is nowhere dense.
We compute the critical Hausdorff dimension and find values significantly different from that found for the AA model, showing that it is a non-universal quantity.
Moreover, we determine accurate values of the critical quasiperiodic amplitude $\Vc$ from the scaling of the inverse participation ratio.
While $\Vc$ depends on the model, we find the universal critical exponent $\nu \simeq 1/3$.

\lettersection{Model and approach}
The single-particle wave functions $\psi(x)$ are found by solving numerically the continuous-space, 1D Schr\"odinger equation:
\begin{equation}\label{eq:Schrodinger}
E \psi (x) = -\frac{\hbar^2}{2m}\frac{d^2 \psi}{d x^2}+V(x) \psi(x),
\end{equation}
using exact diagonalization for Dirichlet absorbing boundary conditions, $\psi_n(0)=\psi_n(L)=0$.
Here, $E$ and $m$ are the particle energy and mass, respectively, $L$ is the system size, and $\hbar$ is the reduced Planck constant.
In the first part of this work, we consider the bichromatic lattice potential
\begin{equation}\label{eq:bichromatic}
V(x) = \frac{V_1}{2}\cos \left(2k_1 x\right)+\frac{V_2}{2}\cos \left(2k_2 x + \varphi\right),
\end{equation}
where the quantities $V_j$ ($j=1,2$) are the amplitudes of two periodic potentials of incommensurate spatial periods $\pi/k_j$ with $k_2/k_1=r$, an irrational number.
The relative phase shift $\varphi$ is essentially irrelevant, except for some values, which induce special symmetries.
In the following, we use $r=(\sqrt{5}-1)/2$ and $\varphi=4$, which avoids such cases.
We then characterize the localization of an eigenstate $\psi$ using the
second-order inverse participation ratio (IPR)~\cite{evers2008},
\begin{equation}\label{eq:IPR}
\IPR = \frac{\int dx\, \vert\psi_n(x)\vert^{4}}{\left(\int dx\, \vert\psi_n(x)\vert^{2} \right)^2}.
\end{equation}
It generally scales as $\IPR \sim 1/L^{\tau}$, with $\tau=1$ for an extended state and $\tau=0$ for a localized state.

\begin{figure}[t!]
\includegraphics[width = 1.01\columnwidth]{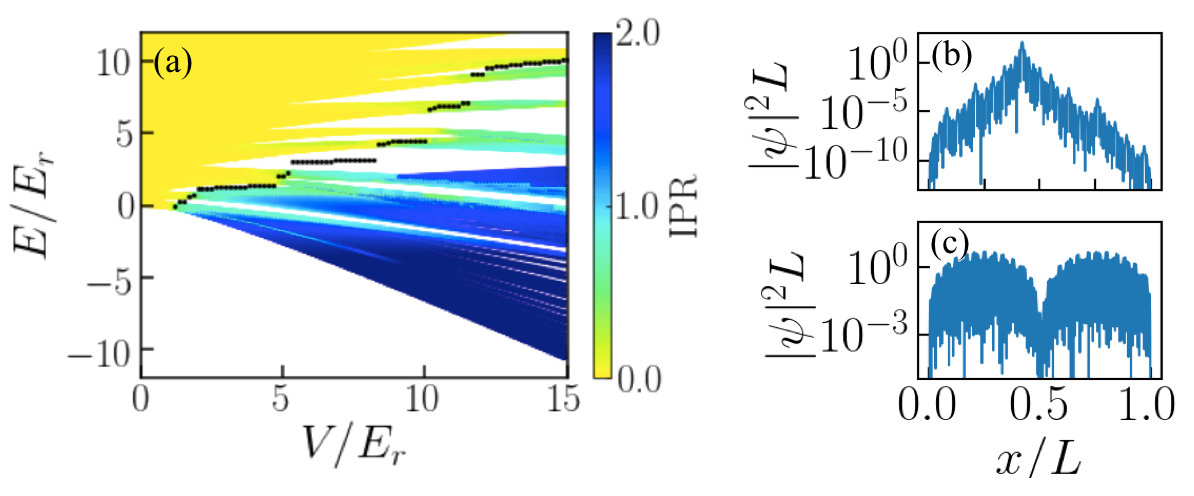}
\vspace{-0.6cm}
\caption{\label{fig:IPReqLatt}
Localization transition for the balanced bichromatic potential, Eq.~(\ref{eq:bichromatic}) with $V_1 = V_2 \equiv V$.
(a)~IPR versus the particle energy $E$ and the lattice amplitude $V$ for the system size $L=100a$.
Localized states correspond to large values of the IPR (blue) and extended to vanishingly small values (yellow).
The ME, found from finite-$L$ scaling analysis of the IPR, is shown as black points.
(b) and (c)~Density profiles of two eigenstates in the localized and extended regimes respectively. Here, the two states correspond to energies right below and right above the ME at $V=2E_r$.
}
 \end{figure}

\lettersection{Mobility edge}
We first focus on the balanced bichromatic lattice, Eq.~(\ref{eq:bichromatic}) with $V_1 = V_2 \equiv V$.
Note that this model cannot be mapped onto the AA model, even for $V \gg \Er$, where  $\Er=\hbar^2k_1^2/2m$ is the recoil energy,
since none of the periodic components of $V(x)$ dominates the other.
Figure~\ref{fig:IPReqLatt}(a) shows the $\IPR$ versus the particle energy $E$ and the potential amplitude $V$
for a large system, $L=100a$ with $a=\pi/k_1$ the spatial period of the first periodic potential.
The results indicate the onset of localization (corresponding to large values of the IPR)
at a low particle energy and high potential amplitude,
consistently with the existence of a $V$-dependent energy ME $\Ec$.
This is confirmed by the behavior of the wave functions, which turn from exponentially localized at low energy~[Fig.~\ref{fig:IPReqLatt}(b)] to extended at high energy~[Fig.~\ref{fig:IPReqLatt}(c)].
These results are characteristic of 1D quasiperiodic models that break the AA self-duality condition~\cite{boers2007,biddle2009,li2017}. The IPR, however, varies smoothly with the particle energy, and is not sufficient to distinguish extended states from states localized on a large scale.

To determine the ME precisely, we perform a systematic finite-size scaling analysis of the IPR and compute the quantity
\begin{equation}\label{eq:derivIPR}
\tau (L) \equiv - d \log\IPR / d\log L.
\end{equation}
For all values of $V$ and $E$, and for large enough system lengths, we find
either $\tau = 0 \pm 0.2$ or $\tau = 1 \pm 0.2$~\cite{note:SupplMat}.
It shows the existence of a sharp localization transition (ME) between localized states ($\tau \simeq 0$) at low energy
and extended states ($\tau \simeq 1$) at high energy.
No intermediate behavior is found in the thermodynamic limit.
The ME $\Ec$ is then accurately determined as the energy of the transition point between the two values.

The results are plotted in Fig.~\ref{fig:IPReqLatt}(a) (black dots).
In all cases, we find that the ME is in an energy gap.
While it is clearly seen for some potential amplitudes (\eg\ for $5.2 \lesssim V \lesssim 8.5$), it is more elusive for some other values (\eg\ for $V/\Er \gtrsim 8.5$), see Fig.~\ref{fig:IPReqLatt}(a).
In the latter case, however, it can be seen by enlarging the figure~\cite{note:SupplMat}.
More fundamentally, it is a direct consequence of the fractal behavior of the energy spectrum, as we discuss now.

\begin{figure}[t!]
\includegraphics[width = 1.0\columnwidth]{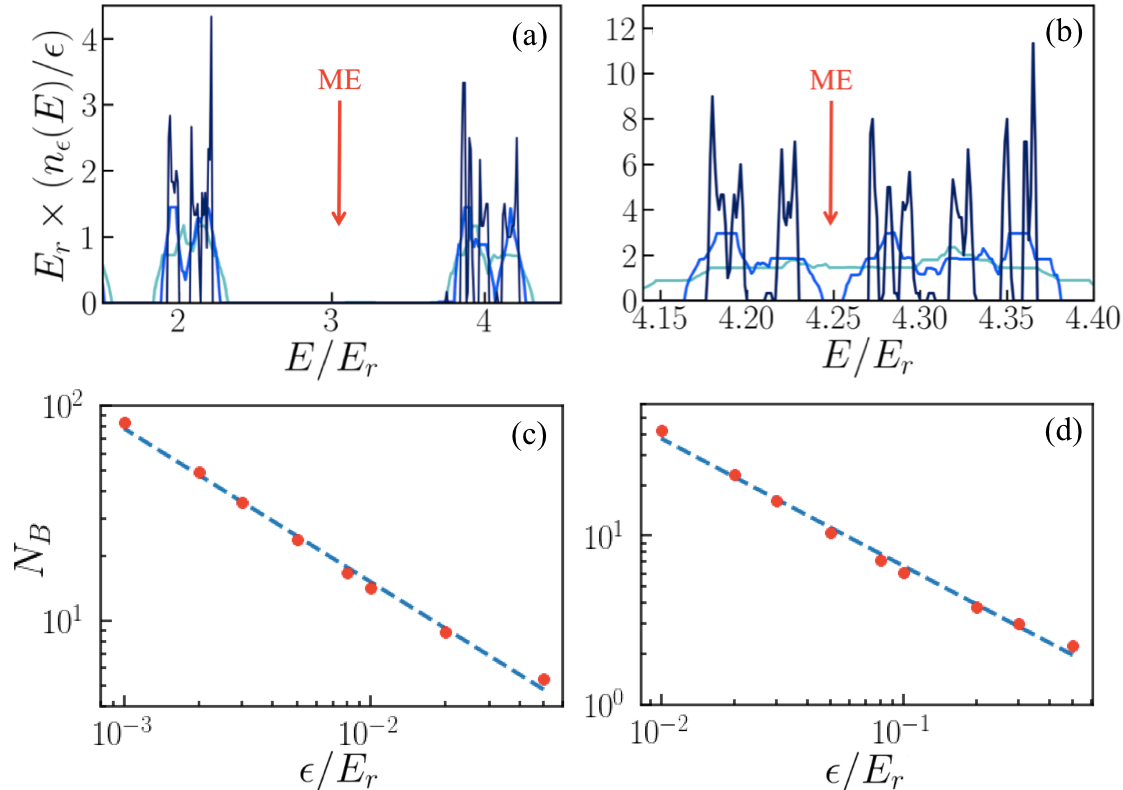}
\caption{\label{fig:DOS}
Fractal behavior of the energy spectrum.
(a) shows $n_\epsilon (E)/\epsilon$ in the vicinity of the ME
at $V=6.0\Er$ for $L=600a$ and $\epsilon/\Er = 0.1$~(light blue), $0.05$~(blue), $0.01$~(dark blue).
(b) shows the same quantity for the ME at $V=8.5\Er$ for $L=1000a$ and $\epsilon/Er=0.1$~(light blue), $0.03$~(blue), $0.003$~(dark blue).
(c) and (d) show the energy-box counting number $\NB$ versus $\epsilon$
for the parameters of (a) and (b), respectively. The linear slopes in log-log scale
are consistent with a fractal behavior, Eq.~(\ref{eq:Hausdorff}) with
$D_H=0.72\pm0.03$ and $D_H=0.76\pm0.03$, respectively.
}
 \end{figure}

\lettersection{Fractality of the energy spectrum}
To characterize the energy spectrum, we first compute the integrated density of states (IDOS) per unit lattice spacing
$n_\epsilon (E)$,
\ie\ the number of eigenstates in the energy range $[E-\epsilon/2,E+\epsilon/2]$,
divided by $L/a$.
Figures.~\ref{fig:DOS}(a) and (b) show the quantity $n_\epsilon (E)/\epsilon$ in the vicinity of the ME for two values of the quasiperiodic amplitude $V$ and several energy resolutions $\epsilon$~\cite{note:DOS}.
For any value of $\epsilon$, the IDOS displays energy bands separated by gaps.
However, when the resolution $\epsilon$ decreases (corresponding to increasingly dark lines on the plots),
new gaps appear inside the bands, while the existing gaps are stable.
It signals that the spectrum is nowhere dense while the gaps are dense in the thermodynamic limit.
In particular, the density of states $\lim_{\epsilon \rightarrow 0_+} n_\epsilon (E)/\epsilon$ is singular.
Moreover, the ME is always found in a gap for a sufficiently resolved spectrum: see Figs.~\ref{fig:DOS}(a) and ~\ref{fig:DOS}(b).
Note that this is not a finite-size effect: For all the results shown here, we have used large enough systems so that each $\epsilon$-resolved band contains at least $10$ - $15$ states. In addition, we have checked that the IDOS is stable against further increasing the system's length~\cite{note:SupplMat}.
The opening of an infinite series of minigaps is characteristic of a fractal behavior.

So far, the fractal character of the energy spectrum of 1D incommensurate systems has been studied for discrete models, such as the Fibonacci chain and the AA model~\cite{hofstadter1976,kohmoto1983,tang1986,kohmoto1987,roscilde2008,tanese2014}.
It was shown that in these cases the spectrum is homeomorphic to a Cantor set.
To study fractality in our continuous model, we use a direct box-counting analysis~\cite{mandelbrot1982,theiler1990}:
We introduce the \textit{energy-box counting number},
\begin{equation}
\NB(\epsilon)=\lim_{q\rightarrow0_+}\int_{E_1}^{E_2}\frac{dE}{\epsilon}\  \big[n_\epsilon(E)\big]^q,
\end{equation}
for some energy range $[E_1,E_2]$.
In the limit $q \rightarrow 0_+$, the quantity $\big[n_\epsilon(E)\big]^q$ approaches $1$ if $n_\epsilon(E) \neq 0$
and $0$ if $n_\epsilon(E) = 0$. Therefore, the quantity $\lim_{q\rightarrow0_+} \big[n_\epsilon(E)\big]^q$ contributes $1$ in the boxes of width $\epsilon$ containing at least one state and vanishes in the empty boxes. 
The sum of these contributions, $\NB(\epsilon)$, counts the minimal number of $\epsilon$-wide boxes necessary to cover all the states within the energy range $[E_1,E_2]$.
The scaling of $\NB(\epsilon)$ versus the energy resolution $\epsilon$,
\begin{equation}\label{eq:Hausdorff}
\NB(\epsilon) \sim \epsilon^{-\DH},
\end{equation}
defines the Hausdorff dimension $\DH$ of the energy spectrum. 
In all considered cases, we found a scaling consistent with Eq.~(\ref{eq:Hausdorff}) with $0<\DH<1$.
This is characteristic of a nontrivial fractal ~\cite{note:DH}.
For instance Figs.~\ref{fig:DOS}(c) and (d) show $\NB$ versus $\epsilon$ in the vicinity of the MEs at $V=6\Er$ and $V=8.5\Er$ for the energy ranges corresponding to Figs.~\ref{fig:DOS}(a) and (b), respectively.
We find a linear scaling in the log-log scale, consistent with Eq.~(\ref{eq:Hausdorff}) and the Hausdorff dimensions
$\DH=0.72\pm0.03$ and $\DH=0.76\pm0.03$, respectively.
Both values are significantly smaller than the geometrical dimension $d=1$.
Therefore, the Lebesgue measure of the energy support vanishes, and the spectrum is nowhere dense in the thermodynamic limit. 

Note that the Hausdorff dimension found above significantly differs from that found in previous work at the critical point of the AA model, $\DH \simeq 0.5$~\cite{tang1986,kohmoto1987}.
We conclude that the spectral Hausdorff dimension is  a nonuniversal quantity.
This is confirmed by further calculations we performed. 
For instance, in the AA limit of our continuous model,
$V_1 \gg V_2,\Er$, we recover $\DH = 0.507 \pm 0.005$ at the critical point.
Conversely, we found $\DH=0.605\pm0.014$ at the critical point of the balanced lattice (see below).

\lettersection{Criticality}

 \begin{figure}[t!]
\includegraphics[width = 0.95\columnwidth]{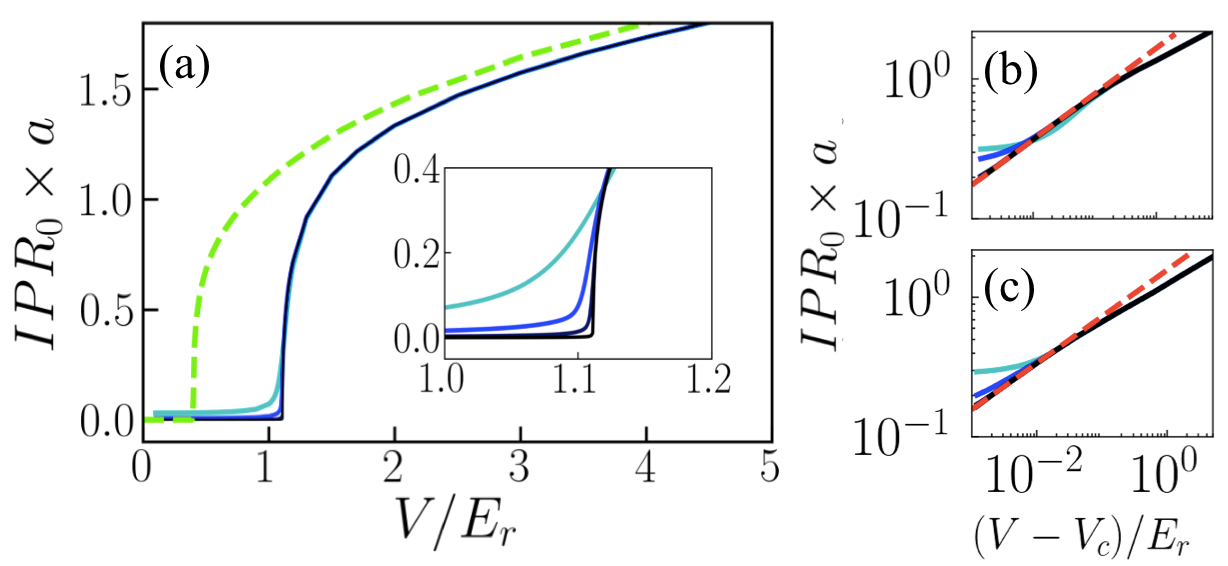}
\caption{\label{fig:critical}
Critical localization behavior.
(a)~Ground-state IPR versus the quasiperiodic amplitude for the balanced bichromatic lattice (solid lines);
Inset: Magnification in the vicinity of the critical point at $\Vc$.
Darker lines correspond to increasing system sizes, $L/a=50$~(light blue), $200$~(blue), $1000$~(dark blue) and $10\ 000$~(black).
The dashed green line corresponds to the trichromatic lattice for $L/a=10\ 000$.
(b), (c)~Ground-state IPR versus $V-\Vc$ in the log-log scale for the bichromatic and trichromatic lattices, respectively.
}
 \end{figure}

We now turn to the critical localization behavior.
As shown in Fig.~\ref{fig:IPReqLatt}(a), a finite ME appears only for a potential amplitude $V$ larger than some critical value $\Vc$, see also Ref.~\cite{biddle2009}.
In Fig.~\ref{fig:critical}(a), we plot the IPR of the ground state ($\IPR_0$) versus $V$.
The transition from the extended phase (vanishingly small IPR) to the localized phase (finite IPR)
gets sharper when the system size increases
and becomes critical in the thermodynamic limit (see darker solid blue lines in the main figure and the inset).
Since the IPR scales as $\IPR_0 \sim 1/L$ in the extended phase and as $\IPR_0 \sim 1$ in the localized phase,
the critical amplitude can be found with a high precision as the fixed point of
$\IPR_0\times\sqrt{La}$ when increasing the system size $L$.
It yields~\cite{note:SupplMat}
\begin{equation}\label{eq:criticalV0}
\Vc/\Er \simeq 1.112\pm 0.002.
\end{equation}
Furthermore, this accurate value of $\Vc$ allows us to determine the critical exponent of the transition. Plotting
$\IPR_0$ versus $V-\Vc$ in log-log scale, we find a clear linear behavior for sufficiently large systems, consistent with the power-law scaling $\IPR_0 \sim (V-V_c)^{\nu}$: see Fig.~\ref{fig:critical}(b).
Fitting the slope, we find the critical exponent
$\nu \simeq 0.327 \pm 0.007$.
Note that for $V \gg \Vc$, $\Er$, the behavior of the IPR changes. The ground state is no longer at criticality and we find the scaling $\IPR_0 \sim V^{\nu^\prime}$ with $\nu^\prime \simeq 0.258\pm0.005$. This is consistent with the exponent $1/4$ expected in the tight-binding limit~\cite{note:SupplMat}.

\lettersection{Other quasi-periodic lattices and universality}

\begin{figure}[t!]
\includegraphics[width = 0.8\columnwidth]{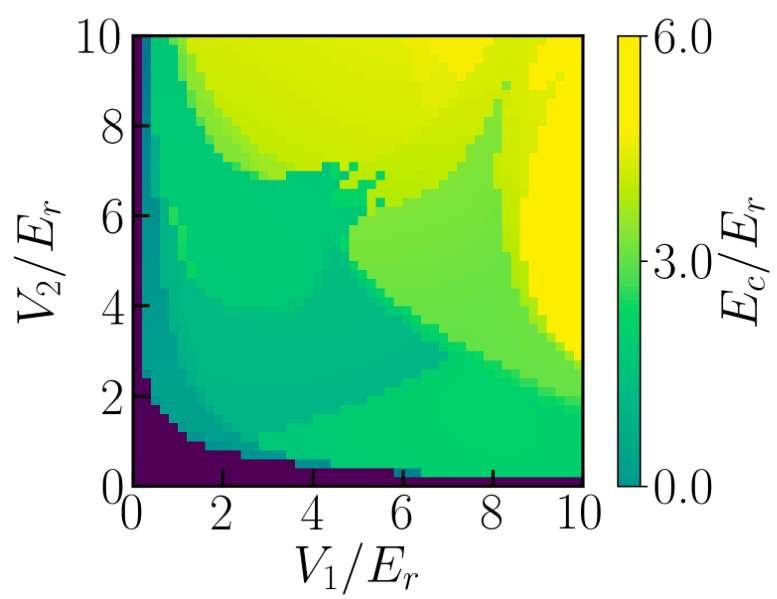}
\caption{\label{fig:V1andV2}
Mobility edge for the imbalanced bichromatic lattice, Eq.~(\ref{eq:bichromatic}) versus the amplitudes $V_1$ and $V_2$.
The dark region indicates the absence of a mobility edge, and its boundary the localization critical line.}
 \end{figure}

We now extend our results to other quasiperiodic models.
We first consider the imbalanced bichromatic lattice, Eq.~(\ref{eq:bichromatic}) with $V_1 \neq V_2$.
In Fig.~\ref{fig:V1andV2}, we plot the ME versus the quasiperiodic amplitudes $V_1$ and $V_2$.
The dark region corresponds to cases where the ME is absent.
Its boundary yields the critical line in the $V_1$-$V_2$ plane.
Note that Fig~\ref{fig:V1andV2} is not symmetric by exchange of $V_1$ and $V_2$ even upon rescaling the energies. This owes to the strong dependence of the model on the incommensurate ratio $r$.
We found that the localization transition is universal, and the critical and fractal properties discussed above for the balanced case
apply irrespectively to the relative amplitudes of the two lattices, \ie\ also for $V_1 \neq V_2$~\cite{note:SupplMat}.
In particular, beyond the critical line, the ME still marks a sharp transition between exponentially localized and extended states, with no intermediate phase.
The energy spectrum is fractal, with $\DH<1$, and thus nowhere dense.
Moreover, for any value of $V_1$
up to values deep in the AA limit ($50\Er$),
we always found $\IPR_0 \sim (V_2-V_{2\textrm{c}})^{\nu}$ with $\nu \simeq 0.33 \pm 0.02$.
The same applies to the discrete AA model~\cite{note:SupplMat}.

It is worth noting that the behavior of the IPR differs from that of the Lyapunov exponent (inverse localization length).
The IPR is dominated by the core of the wave function and characterizes, for instance, the
short-range interaction energy of two particles in a localized state~\cite{lugan2007a}.
In contrast, the Lyapunov exponent $\gamma$ characterizes the exponential tails of the wave functions,
$\psi(x) \sim \exp(-\gamma \vert x \vert)$,
and it is insensitive to the core.
For nonpurely exponential wave functions, which appear in our model (see for instance Fig.~\ref{fig:IPReqLatt}(b) and Ref.~\cite{note:SupplMat}), these two quantities are not proportional.
For instance, in the AA model, one has $\gamma \sim \ln(\Delta/2J)$,
and, at the critical point $\Delta=2J$, one finds $\gamma \sim \Delta/2J \sim (V_2-V_{2\textrm{c}})^{\beta}$
with the Lyapunov critical exponent $\beta=1$. This value differs from the IPR critical exponent-$\nu \simeq 1/3$-found above.

We also considered the trichromatic lattice
\begin{equation}\label{eq:trichromatic}
V(x) = \frac{V}{2} \big[\cos \left(2k_1 x\right)+\cos \left(2k_2 x + \varphi\right)+\cos \left(2k_3 x + \varphi^{\prime}\right) \big],
\end{equation}
with $k_3/k_2=k_2/k_1=r$, so that the three lattice spacings are incommensurate to each other [note that $k_3/k_1=r^2=(3-\sqrt{5})/2$ is an irrational number].
Performing the same analysis as for the other models, we recover the same universal features. In particular, the energy spectrum is fractal and nowhere dense, and the mobility edge is always in a gap. We find a finite critical amplitude $\Vc$ and the critical behavior $\IPR_0 \sim (V-\Vc)^{\nu}$ with $\nu \simeq 0.327 \pm 0.007$; see Fig~\ref{fig:critical}(c). The only significant difference is that the critical point for the trichromatic lattice, $\Vc/\Er\simeq0.400\pm 0.005$, is smaller than for the bichromatic lattice; see Fig.~\ref{fig:critical}(a).
In particular, the standard deviation of the potential, $\Delta V$, is a factor about $2.27$ smaller at the critical point.
This is consistent with the intuitive expectation that it should vanish in the disordered case corresponding to an infinite series of cosine components with random phases~\cite{lifshits1988,beenakker1997}.

\lettersection{Conclusion}
In summary, we have studied the critical and fractal behavior for single particles in quasiperiodic potentials.
Our results shed light on models that have become pivotal for Anderson~\cite{li2017,luschen2018} and many-body~\cite{kohlert2019} localization.
We found that the ME is always in a gap and separates localized and extended states, with no intermediate phase.
We related this behavior to the fractality of the energy spectrum and found that the Hausdorff dimension is always smaller than unity but nonuniversal.
In contrast, we found the critical behaviour $\IPR_0 \sim (V-\Vc)^{\nu}$
with the  universal exponent $\nu \simeq 1/3$.
These predictions may be confirmed in experiments similar to Ref.~\cite{luschen2018} using energy-resolved state selection~\cite{pezze2011a,volchkov2018,richard2019}. In parallel to further theoretical studies, they may help answer questions our results call. For instance, it would be interesting to determine the physical origin of the critical exponent $\nu$ and extend our study to higher dimensions. Another important avenue would be to extend it to interacting models in connection to many-body localization.

Our results may also pave the way to the observation of the still elusive Bose-glass phase.
So far ultracold-atom experiments have been performed in the AA limit, the energy scale of which is the tunneling energy $J$~\cite{roux2008}. The latter is exponentially small in the main lattice amplitude and of the order of the temperature. It suppresses coherence, and significantly alters superfluid-insulator transitions~\cite{derrico2014,gori2016}.
In shallow quasiperiodic potentials, the energy scale is, instead, the recoil energy $\Er$,
which is much higher than the temperature.
Temperature effects should thus be negligible.
For strong interactions, the 1D Bose gas can be mapped onto an ideal Fermi gas and
the Bose-glass transition is directly given by the ME we computed here.
It would be interesting to determine how the transition evolves for weak interactions.

\acknowledgments
We thank David Clément and Thierry Giamarchi for fruitful discussions. This research was supported by the
European Commission FET-Proactive QUIC (H2020 Grant No.~641122) and the Paris region DIM-SIRTEQ.
This work was performed using HPC resources from GENCI-CINES (Grant No.~2018-A0050510300).
We thank the CPHT computer team for valuable support.


 \renewcommand{\theequation}{S\arabic{equation}}
 \setcounter{equation}{0}
 \renewcommand{\thefigure}{S\arabic{figure}}
 \setcounter{figure}{0}
 \renewcommand{\thesection}{S\arabic{section}}
 \setcounter{section}{0}
 \onecolumngrid  
     
 
 \newpage

{\center \bf \large Supplemental Material for \\}
{\center \bf \large \letitre \\ \vspace*{1.cm}
}

This supplemental material gives details about
the finite-size scaling analyses used for the mobility edge, the integrated density of states, and the critical point (Sec.~\ref{sec:scaling}), 
the sketch for the mobility edge and critical point (Sec.~\ref{sec:sketch}),
the critical and fractal behaviours for the imbalanced case (Sec.~\ref{sec:imbalanced}),
as well as about the tight-binding and AA limits (Sec.~\ref{sec:AA-model}).

\section{Finite-size scaling analyses}\label{sec:scaling}
Here we discuss the finite-size scaling analyses performed for the determination of
the mobility edge~(Sec.~\ref{sec:scaling.IPR}),
the fractal character of the energy spectrum~(Sec.~\ref{sec:scaling.IDOS}),
and the critical potential amplitude~(Sec.~\ref{sec:critical}).

\subsection{Inverse participation ratio}\label{sec:scaling.IPR}
\begin{figure}[h!]
\includegraphics[width = 0.76\columnwidth]{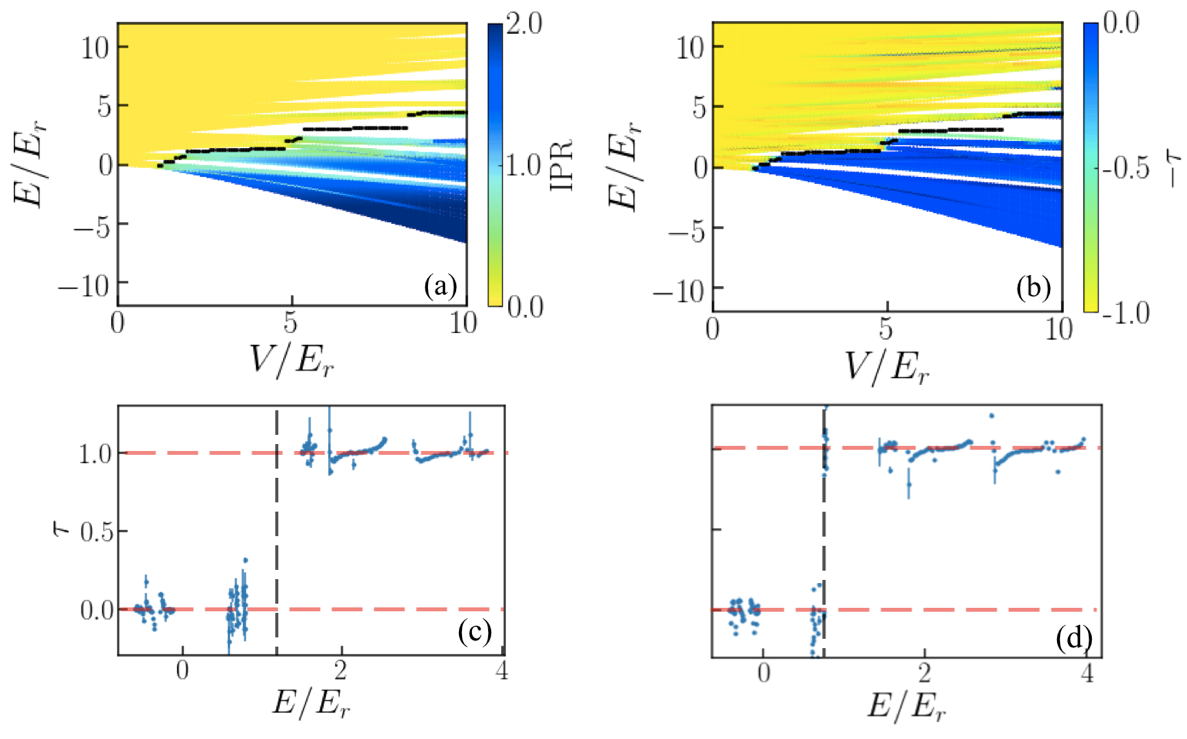}
\caption{\label{fig:finite-size}
Accurate determination of the energy mobility edge $\Ec$ for the balanced bichromatic lattice.
Panel~(a) shows the IPR versus the particle energy $E$ and the lattice amplitude $V$ for the system size $L=100a$
[reproduced from Fig.~1(a) of the main paper].
Panel~(b) shows the exponent $\tau$ versus $E$ and $V$, as found from finite-size scaling analysis of data computed for various system sizes.
Panels~(c) and (d) are cuts of panel~(b) at $V=2\Er$ and $V=1.7\Er$, respectively.
The system size ranges from $L=50a$ to $L=800a$ for most of the points.
When the ME lies in a very small gap, as for panel~(d) for instance we use larger system sizes, typically up to $L=1000a$.
}
 \end{figure}

The scaling analysis of the inverse participation ratio (IPR) is performed as follows.
For each value of the quasi-periodic amplitude $V$, we diagonalize the Hamiltonian for a series of system sizes $L$,
typically ranging from $50a$ to $800a$ ($a$ is the spacing of the first lattice).
For any state, we find that the IPR scales as $\IPR \sim L^{\tau}$, with either $\tau=0 \pm 0.2$ or $\tau=1\pm0.2$.
Therefore, in contrast to the IPR at a given  system length, which varies smoothly [see Fig.~1(a) of the main paper, also reproduced on Fig.~\ref{fig:finite-size}(a)], the exponent $\tau$ shows a sharp transition from localized states (corresponding to  $\tau \simeq 0$) to extended states (corresponding to $\tau\simeq1$), see Fig.~\ref{fig:finite-size}(b) as well as Fig.~\ref{fig:finite-size}(c) and (d) for two typical cuts at fixed values of the quasi-periodic amplitude.
The mobility edge (ME) $\Ec$ is then determined as the transition point between the values of $\tau$, see black points on Fig.~\ref{fig:finite-size}(b). More precisely, as discussed in the main paper, the ME is always in an energy gap and we define $\Ec$ as the average energy of the last localized state and the first extended state, see dashed black lines on Figs.~\ref{fig:finite-size}(c) and (d), corresponding to a large and small gap, respectively.

\subsection{Integrated density of states}\label{sec:scaling.IDOS}
\begin{figure}[h!]
\bigskip
\includegraphics[width = 0.9\columnwidth]{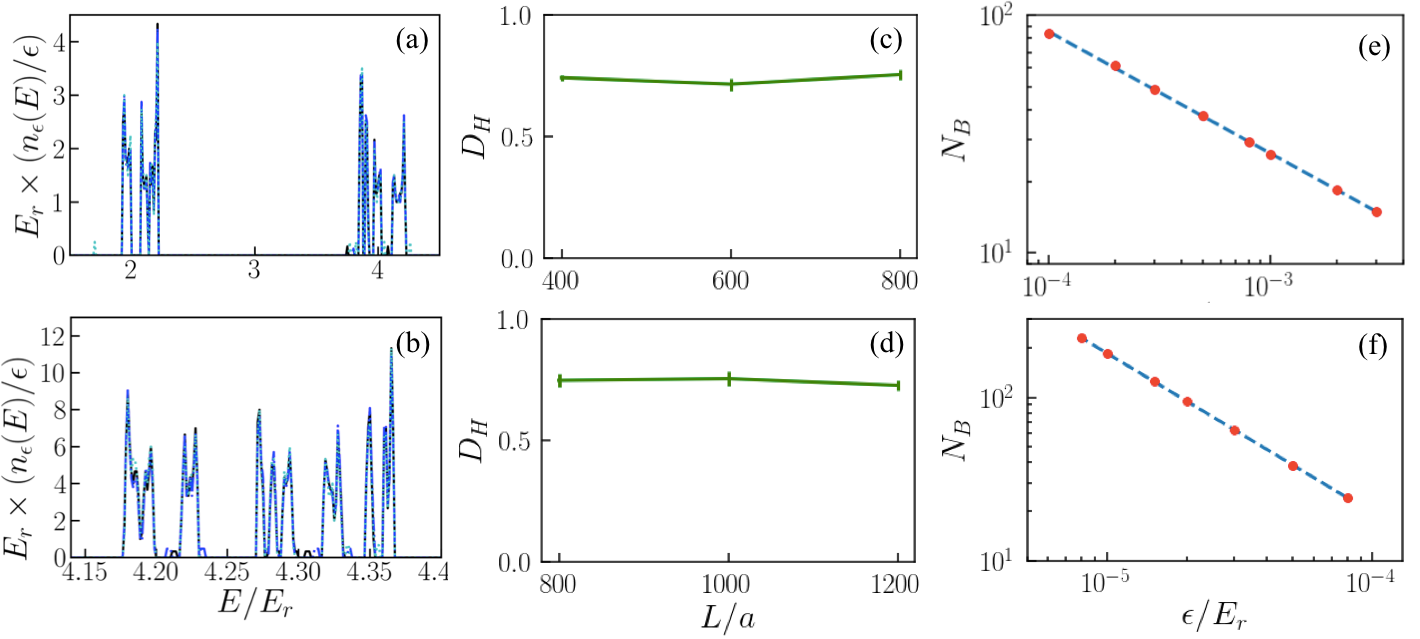}
\caption{\label{fig:idos-size}
Finite-size scaling analysis for the IDOS.
The panels (a) and (b) reproduce the IDOS divided by the energy resolution, $n_{\epsilon}(E)/\epsilon$, shown on Fig.~2 of the main paper for the smallest considered energy resolutions $\epsilon$ but for various values of the length $L$.
(a)~IDOS in the vicinity of the ME at $V=6\Er$ for $\epsilon=0.01\Er$
and $L=400a$~(dotted light blue line), $600a$~(dashed blue line), $800a$~(solid black line).
(b)~Same as panel~(a) for $V=8.5\Er$, $\epsilon=0.003\Er$,
and $L=800a$~(dotted light blue line), $1000a$~(dashed blue line), $1200a$~(solid black line). 
(c) and (d) Hausdorff dimension $\DH$ calculated for the various system sizes used for the panels~(a) and (b).
(e) and (f) Energy-box counting number $\NB$ versus $\epsilon$
for (e)~the continuous bichromatic model in the Aubry-Andr\'e limit, $V_1=10\Er$ and $V_2=0.09\Er$, at criticality
and (f)~the commensurate bichromatic lattice, $r=2/3$ and $V_1=V_2=6\Er$.
}
 \end{figure}

To show that the opening of mini gaps in the energy spectrum is not due to finite-size effects, we have computed the integrated density of states (IDOS) for various system lengths.
Figures~\ref{fig:idos-size}(a) and (b) reproduce the IDOS shown on Fig.~2 of the main paper for the smallest considered energy resolutions $\epsilon$ and various values of the length $L$.
The results corresponding to the different system lengths are indistinguishible. 
Moreover, we have computed the Hausdorff dimensions in both cases for the different lengths, see Figs.~\ref{fig:idos-size}(c) and (d). The behaviours of $\DH$ do not show significant variations with the system size.
These results allow us to rule out finite-size effects.

To further confirm the fidelity of our approach, we have performed two additional checks.
On the one hand, we have computed the Hausdorff dimension of the first band of the first lattice for the continuous model in the Aubry-Andr\'e limit ($V_1 \gg V_2, \Er$).
We find a clear fractal behaviour of the energy-box number, $\NB \sim \epsilon^{-\DH}$ with $\DH=0.51\pm0.01$,
see Fig.~\ref{fig:idos-size}(e). It is in excellent agreement with the Hausdorff dimension found by another method in the discrete Aubry-Andr\'e model, $\DH \simeq 0.5$~\cite{tang1986,kohmoto1987}.
On the other hand, we have reproduced the same calculation as for the case corresponding to Fig.~\ref{fig:idos-size}(a), \ie\ $V_1=V_2=6\Er$, but with the commensurate filling $r=2/3$.
It corresponds to a periodic system and a regular spectrum with $\DH=1$ is expected. The result is shown on Fig.~\ref{fig:idos-size}(f) and we find $\DH=0.98\pm0.01$, in excellent agreement with this prediction.
These results further validate our approach to determine the fractal dimension of the energy spectrum.

\subsection{Determination of the critical potential}\label{sec:critical}
\begin{figure}[h!]
\subfigure{ \includegraphics[width=2.0in]{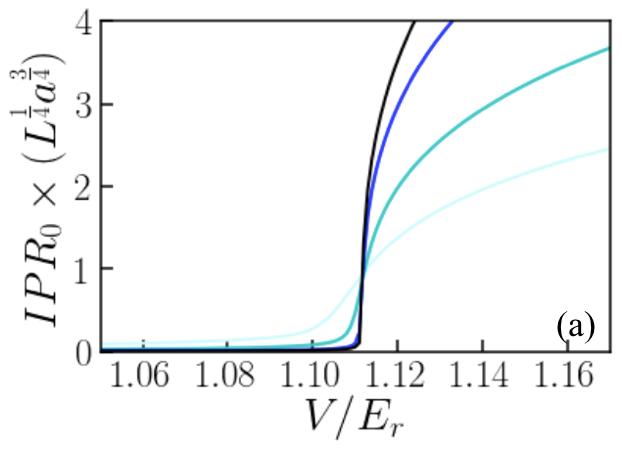}} 
  \hspace{0.05in} 
 \subfigure{ \includegraphics[width=2.0in]{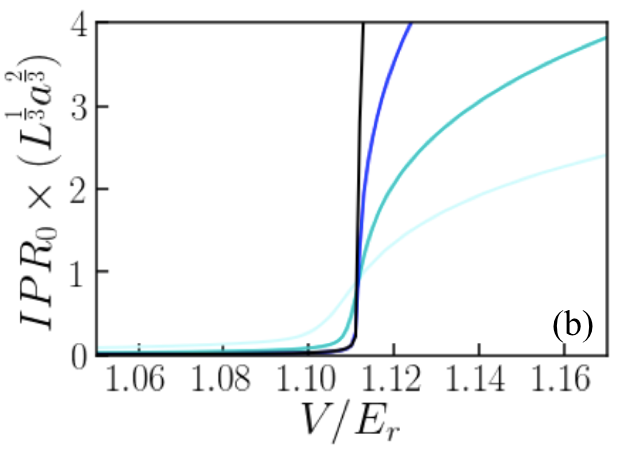}} 
  \hspace{0.05in} 
  \subfigure{ \includegraphics[width=2.1in]{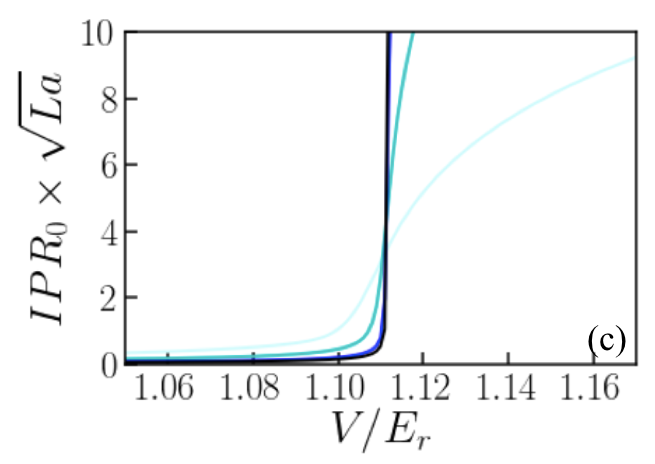}} 
  \hspace{0.05in} 
  \subfigure{ \includegraphics[width=2.0in]{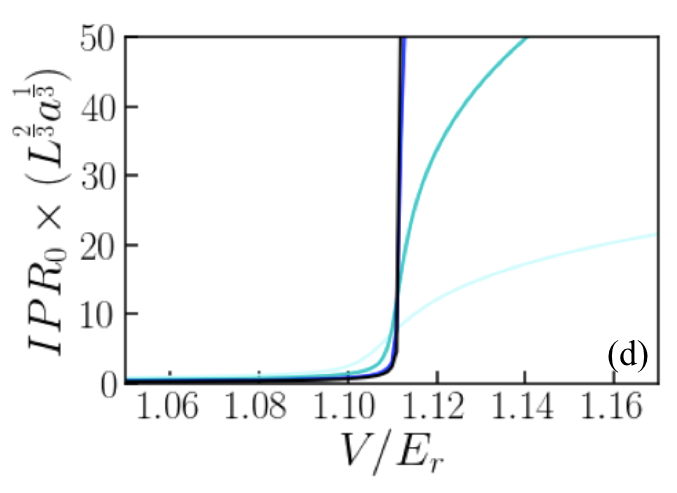}} 
  \hspace{0.05in} 
  \subfigure{ \includegraphics[width=2.1in]{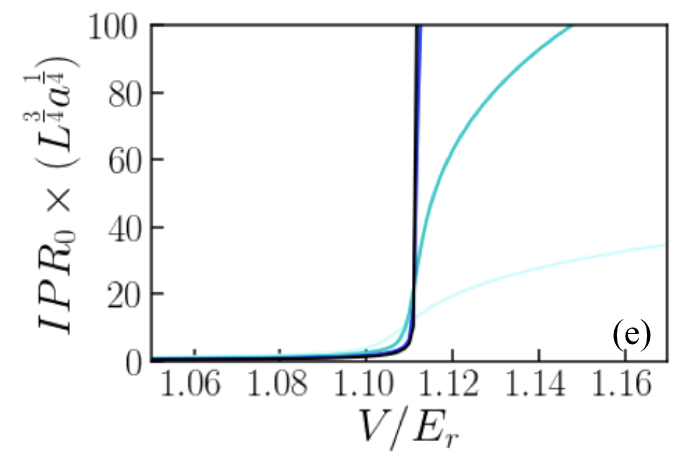}} 
  \hspace{0.05in} 
\caption{\label{response2}
Plots of the quantity $\IPR_0\times L^{\alpha} \times a^{1-\alpha}$ as a function of the potential amplitude $V$ for the balanced bichromatic lattice and different system lengths.
Darker lines correspond to increasing system sizes, $L/a=200$~(light blue), $1000$~(blue), $5000$~(dark blue), $10000$~(black).}
\end{figure}

For all the cases discussed in the main paper, the critical potential amplitude $\Vc$ is determined by plotting the quantity $\IPR_0 \times L^\alpha a^{1-\alpha}$ versus $V$ with $\alpha=1/2$. More generally, one may use any value $0<\alpha<1$, as discussed here.
Then, for a localized state, the quantity $\IPR_0 \times L^\alpha a^{1-\alpha}$ increases with $L$, while for an extended state it decreases with $L$. The turning point between these two opposite behaviours yields an accurate value of $\Vc$.
Figure~\ref{response2} shows this approach for the balanced bichromatic lattice.
For large enough systems and any of the considered values of $\alpha$, the curves corresponding to different lengths cross each other at almost the same value of $V/\Er$. For the various values of $\alpha$ considered here, we find the following estimates:
\begin{table}[h!]
\centering
\begin{tabular}{|c|c|c|c|c|c|}
    \hline
    $\alpha$ & 1/4 & 1/3 & 1/2 & 2/3 & 3/4 \\ \hline
    $\Vc/\Er$ & 1.113 & 1.111 & 1.112 & 1.111 & 1.110 \\ \hline
    accuracy & 0.004 & 0.002 & 0.002 & 0.002 & 0.004 \\ \hline
  \end{tabular}
\end{table}

\noindent
All the results agree within the errorbars. As expected, the most accurate result is found for $\alpha=1/2$, which maximally discriminates the localized and extended states. It yields the value $\Vc/\Er \simeq 1.112\pm0.002$~[Eq.~(7) of the main paper].

\section{Mobility edge and critical point}\label{sec:sketch}
\begin{figure}[h!]
\includegraphics[width = 0.8\columnwidth]{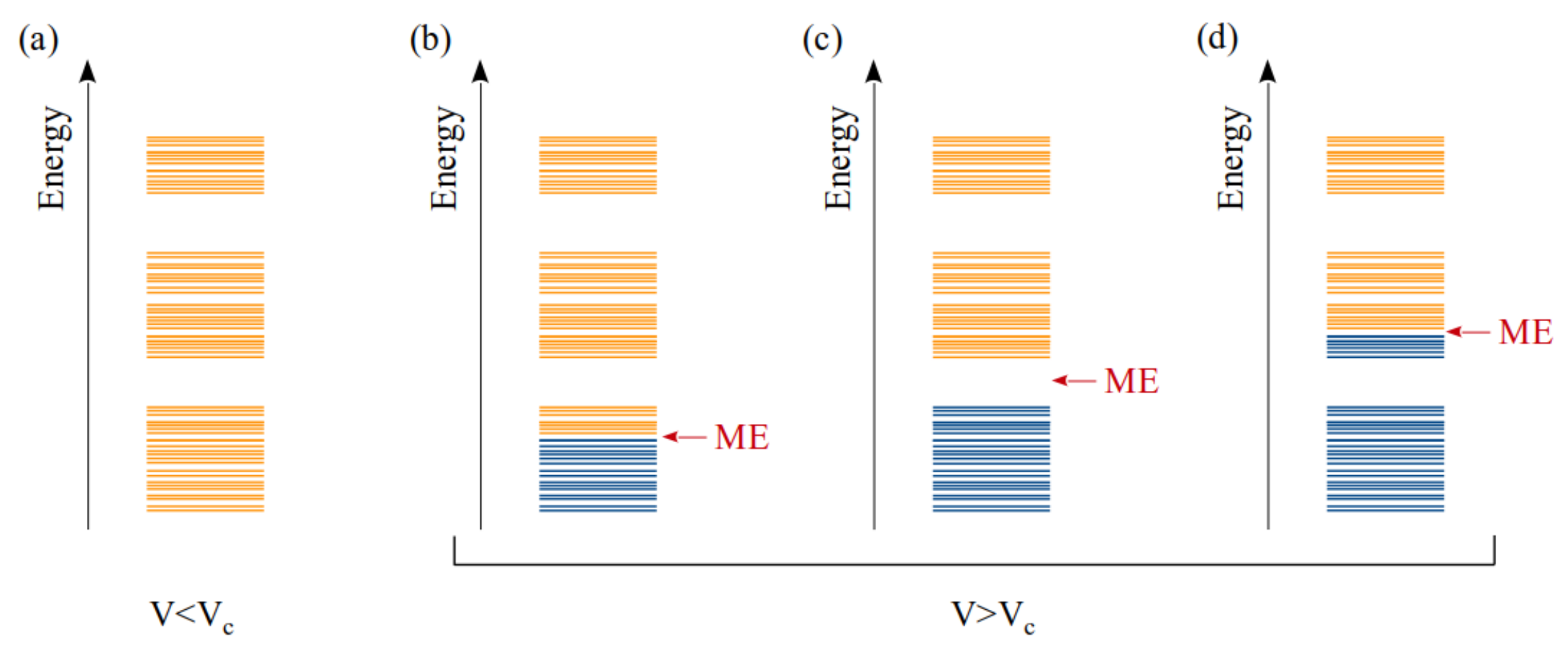}
\caption{\label{regimes}
Localization in a shallow quasi-periodic lattice.
The extended states are shown in orange and the localized states in blue.
(a)~Case $V<\Vc$ where all the states are extended.
(b), (c), and (d)~Three examples in the case $V>\Vc$, with different values of the mobility edge $\Ec$. The potential $V$ increases from (b) to (d).
}
 \end{figure}

Figure~\ref{regimes} shows a simplified sketch of the energy spectrum in various cases. 
The extended states are represented in orange and the localized states in blue.
For $V < \Vc$, all the states are extended and there is no ME, see Fig.~\ref{regimes}(a).
For $V > \Vc$, the ME is finite: The lowest-energy states are localized and the highest-energy states are extended, see Figs.~\ref{regimes}(b), (c), and (d). As discussed in the main text, the ME is always found in a gap, be it small~[cases~(b) and (d)] or large~[case~(c)].
The "intermediate phase" discussed in Refs.~\cite{li2017,luschen2018} corresponds to the case~(b) where the ME is below the lowest large gap.
In the imbalanced case, it roughly corresponds to the lowest band of the strongest lattice.
For $V$ slighly above the critical point $\Vc$, this portion of the spectrum contains both localized states at low energy and extended states at high energy.
These states are, however, well separated by an energy gap near by the ME.
Here the gap is rather small but this case is formally not different from the case of Fig.~\ref{regimes}(c) where the ME lies in a large gap.
While distinguishing the two cases may be useful in practice, there is no formal difference
--~and no phase transition~-- between the cases (b) and (c). In both, the ME is finite and lies in a gap.

\section{Critical and fractal behaviours for the imbalanced case}\label{sec:imbalanced}

To show that the critical and fractal properties discussed in the main text for the balanced lattice ($V_1=V_2$) extend to the imbalanced lattice, irrespective of the relative amplitudes of $V_1$ and $V_2$, we consider here the case with $V_1=8\Er$ and scan the value of $V_2$. This case was also considered in Refs.~\cite{li2017,luschen2018}.

\begin{figure}[h!]
\includegraphics[width = 1.0\columnwidth]{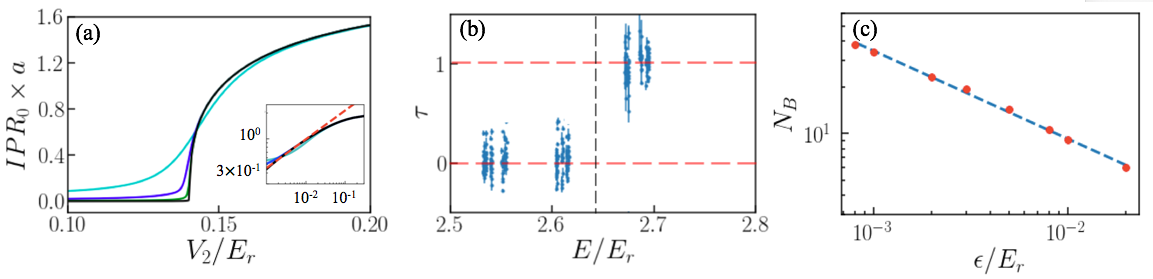}
\caption{\label{imbalance}
Critical and fractal behaviour for the imbalanced bichromatic lattice.
Here we use $V_1=8.0\Er$ and $r=(\sqrt{5}-1)/2$.
(a)~IPR of the ground state, $\IPR_0$, versus the amplitude of the second lattice $V_2$. Darker lines correspond to increasing system sizes, $L/a=50$~(light blue), $200$~(blue), $1000$~(dark blue), $10000$~(black).
Inset: Same data versus $V_2-V_{2\textrm{c}}$ in log-log scale, confirming the critical behaviour $\IPR_0 \sim (V_2-V_{2\textrm{c}})^\nu$, with $\nu \simeq 0.33 \pm 0.01$ (dashed red line).
(b)~Scaling exponent $\tau$ versus the energy $E$ as found from fits as $\IPR \sim 1/L^\tau$, for the specific case $V_2=0.15\Er>V_{2\textrm{c}}$. The black dashed line marks the ME.
(c)~Energy box-counting number $\NB$ as a function of the energy resolution $\epsilon$, in the energy window corresponding to panel~(b).
}
 \end{figure}

Here, the ME is found at $V_{2\textrm{c}}/\Er \simeq 0.140 \pm 0.005$, see Fig.~\ref{imbalance}(a). Using the same analysis as for the balanced lattice (see main text), we find the critical behaviour
\begin{equation}
\IPR_0 \sim (V_2-V_{2\textrm{c}})^\nu
\qquad \textrm{with} \qquad
\nu \simeq 0.33 \pm 0.01,
\end{equation}
see inset of Fig.~\ref{imbalance}(a).
This value of the critical exponent $\nu$ is very close to that found for the balanced case.

We now study the ME for $V_2 = 0.15\Er > V_{2\textrm{c}}$ and perform the finite-size analysis. We compute the IPR versus the energy $E$ for various system sizes. For any $E$, we find the scaling $\IPR \sim 1/L^\tau$ with either $\tau \simeq 0$ or $\tau \simeq 1$, see Fig.~\ref{imbalance}(b).
A sharp jump from $\tau \simeq 0$ to $\tau \simeq 1$ marks the ME, here found at $\Ec \simeq 2.64$. As shown by Fig.~\ref{imbalance}(b), the ME is in a gap. It confirms that the transition is sharp, between a localized phase and an extended phase, with no intermediate phase. 

This is further confirmed by the box-counting analysis of the energy spectrum. 
We compute the energy box-counting number $\NB$ as a function of energy resolution $\epsilon$ for the same energy range as in Fig.~\ref{imbalance}(b).
The result, shown in Fig.~\ref{imbalance}(c), yields the characteristic fractal behaviour $\NB \sim \epsilon^{-\DH}$ with $\DH=0.57\pm0.02$. The factal dimension $\DH$ is significantly smaller than unity. It further confirms that the spectrum is nowhere dense and, in particular, that the ME can only be in a gap.

\section{Tight-binding limit and Aubry-Andr\'e  model}\label{sec:AA-model}

\subsection{Inverse participation ratio of the ground state in deep lattices}
\begin{figure}[h!]
\includegraphics[width = 0.65\columnwidth]{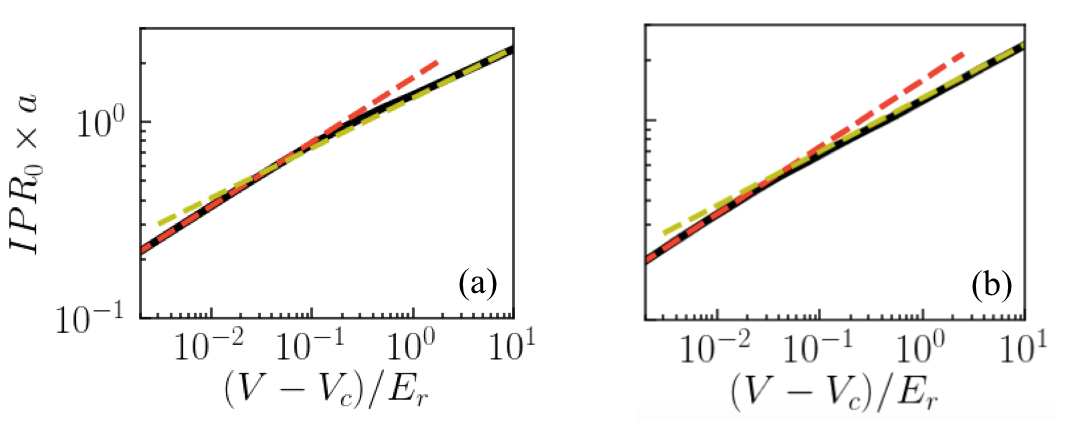}
\caption{\label{fig:ipr0-largev}
Crossover of ground-state IPR (solid black line) from the critical behaviour $\IPR_0 \sim (V-\Vc)^\nu$ with $\nu \simeq 1/3$ (dashed red line) at criticality to the asymptotic behaviour $\IPR_0 \sim V^{\nu'}$ with $\nu' \simeq 1/4$ for $V \gg \Vc, \Er$ (dashed yellow line). The panels (a) and (b) correspond to the bichromatic and trichromatic cases, respectively.
}
 \end{figure}

The tight-binding limit is obtained when at least one of the potential amplitudes exceeds the recoil energy,
$V_j \gg \Er$. In this case, the local potential minima support bound states and, for the quasi-periodic potential, the tunneling is suppressed.
Within a harmonic approximation of the potential minima, we find the frequency
$\omega \propto \sqrt{V\Er}/\hbar$. The ground state and, more generally, the lowest energy eigenstates of the quasi-periodic potential are nearly Gaussian functions of width $\ell = \sqrt{\hbar/m\omega}$ and  centered at the bottom of diffrent potential minima.
Then, the IPR of the ground state scales as $\IPR_0 \sim 1/\ell$, \ie\
\begin{equation}
\IPR_0 \sim V^{\nu'}
\qquad \textrm{with} \qquad 
\nu'=1/4.
\end{equation}
This exponent significantly differs from the critical exponent $\nu \simeq 1/3$ found at the critical point $\Vc$.
In Fig.~\ref{fig:ipr0-largev}, we show how $\IPR_0$ crosses over from the critical behaviour $\IPR_0 \sim (V-\Vc)^\nu$ with $\nu \simeq 1/3$ for $V \gtrsim \Vc$ to the asymptotic behaviour $\IPR_0 \sim V^{\nu'}$ with $\nu' \simeq 1/4$ for $V \gg \Vc, \Er$.

\subsection{Aubry-Andr\'e model}
The Aubry-Andr\'e (AA) limit of our continuous bichromatic lattice is found in the tight-binding limit of one of the lattices (say lattice $1$) while the other lattice (say lattice $2$) is weak. Our model can then be mapped onto the discrete, Aubry-Andr\'e model,
\begin{equation}\label{eq:AAmodelH}
\hat{H}_\textrm{\tiny AA} = -J \sum_{\langle i, j \rangle} \left( \hat{a}_i^\dagger \hat{a}_j + \textrm{H.c.} \right) + \Delta \sum_i \cos(2\pi r i + \varphi) \hat{a}_i^\dagger \hat{a}_i,
\end{equation}
where $\hat{a}_i$ is the annihilation operator of a particle in the lattice site $i$ (located at the position $x_i = a \times i$),
$J$ is the tunneling energy associated to lattice $1$ and $\Delta$ is the quasi-periodic amplitude induced by lattice~2.
The AA parameters in Eq.~(\ref{eq:AAmodelH}) are given by
\begin{equation}\label{eq:AAmodelHparam1}
J \simeq \frac{4\Er}{\sqrt{\pi}}\left(\frac{V_1}{\Er}\right)^{3/4}\exp\left(-2\sqrt{\frac{V_1}{\Er}}\,\,\right)
\qquad \textrm{with} \qquad 
\Delta \simeq \frac{V_2}{2} \exp\left(-r^2 \sqrt{\frac{\Er}{V_1}}\,\,\right),
\end{equation}
see for instance Refs.~\cite{damski2003,biddle2010}.

\begin{figure}[t!]
\includegraphics[width = 0.95\columnwidth]{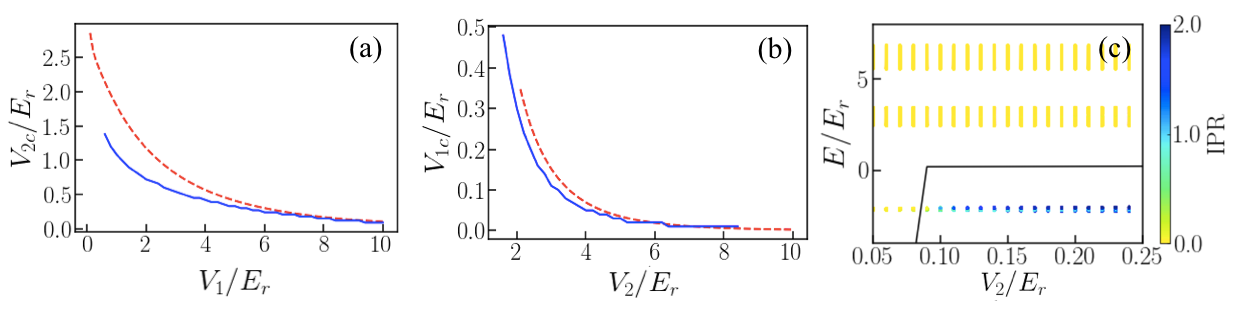}
\caption{\label{fig:AAcheck}
Comparison between the critical point found in the continuous bichromatic model (solid blue lines) and the discrete Aubry-Andr\'e model (dashed red lines).
Panel~(a) corresponds to the tight-binding regime for lattice~1 and panel~(b) to  tight-binding regime for lattice~2, respectively.
Panel (c) shows the lowest part of the energy spectrum for $V_1=10\Er$, as a function of $V_2$.
The color scale encodes the IPR, corresponding to localized (blue) and extended (yellow) states.
The ME is shown as the solid black line.
}
 \end{figure}

On Fig.~\ref{fig:AAcheck}(a), we plot the critical potential of the second lattice, $V_{2\textrm{c}}$, versus the amplitude of the first lattice, $V_1$, for the imbalanced bichromatic lattice.
We find excellent agreement between the results founds for the continuous model (solid blue lines) and the prediction of the discrete AA model, corresponding to $\Delta=2J$~\cite{aubry1980} (dashed red lines), for $V_1 \gtrsim 8\Er$.

Figure~\ref{fig:AAcheck}(b) shows the same comparision in the opposite situation where lattice~2 is in the tight-binding regime and lattice~1 is weak. In this case, the AA parameters are changed into
\begin{equation}
J \simeq \frac{4\Er r^{1/2}}{\sqrt{\pi}}\left(\frac{V_2}{\Er}\right)^{3/4}\exp\left(-2r^{-1}\sqrt{\frac{V_2}{\Er}}\,\,\right)
\qquad \textrm{with} \qquad 
\Delta \simeq \frac{V_1}{2} \exp\left(-r^{-1} \sqrt{\frac{\Er}{V_2}}\,\,\right).
\end{equation}
Then, the critical potential $V_{1\textrm{c}}$ found in the continuous model approaches the AA prediction for $V_2 \gtrsim 6\Er$.

In Fig.~\ref{fig:AAcheck}(c), we show the lowest part of the energy spectrum in the AA limit of the continuous bichromatic model, $V_1=10\Er \gg \Er$, as a function of the amplitude of the second lattice, $V_2$.
The structure of the spectrum is reminiscent of the band spectrum of the dominant lattice, and we refer to the visible states clusters as the first bands of the main lattice. The fractal structure of the spectrum induced by the sub-dominant lattice is not visible here.
The color scale encodes the IPR, corresponding to localized (blue) and extended (yellow) states. The solid black line shows the mobility edge as found from a cut of Fig.~4 of the main paper at the fixed value of $V_1=10\Er$.
For vanishingly small values of $V_2$, there is no ME and all the states are extended. When increasing the value of $V_2$, the ME sharply jumps to a value in the first band gap of the main lattice.
Then, all the states of the first band of the main lattice become localized.
The critical point is found at $V_2 \simeq 0.09\Er$. Using the formulas in Eq.~(\ref{eq:AAmodelHparam1}), we find that it corresponds to $\Delta / 2J \simeq 1.04$, in excellent agreement with the prediction of the discrete AA model~\cite{aubry1980}.
Note that the states of the second and third bands of the main lattice are still extended. They become localized at a higher value of $V_2$, see Fig.~4 of the main paper.

\begin{figure}[h!]
\includegraphics[width = 0.6\columnwidth]{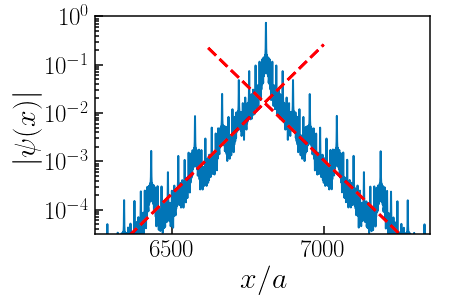}
\caption{\label{fig:Mq}
Ground-state wavefunction of the Aubry-Andr\'e model for $\Delta/J=2.05$ (solid blue line) together with exponential fits (dashed red lines) on the left-hand and right-hand sides of the localization center.
}
 \end{figure}

Finally, we illustrate here the difference between the IPR and the Lyapunov exponent as probes of localization. On Fig.~\ref{fig:Mq}, we plot the ground-state wavefunction for the AA model slightly above the critical point, namely $\Delta/J = 2.05$.
The wavefunction shows a clear exponential localization in the wings. The dahsed red lines are the fitted exponential function
$\vert\psi\vert \propto e^{-\gamma|x-x_0|}$ with the localization center $x_0$ and the Lyapunov exponent $\gamma$ as fitting parameters.
It yields $\gamma=6.2\times10^{-3}\pm6.85\times10^{-5}$.
However, the wavefunction is not a pure exponential function. In particular, it shows a core about one order of magnitude larger than the exponential fit at the localization center $x_0$. This core dominates the IPR.
For instance, restricting the wavefunction to the range $[x_1,x_2]$ such that $\psi(x_1)=\psi(x_2)=0.01\psi(x_0)$, we find that the IPR is $99.5\%$ of the value found for the full wavefunction. Therefore, the IPR is independent of the exponential behaviour of the tails, and the IPR and Lyapunov exponent yield different pieces of information about localization.
In particular, they are characterized by different critical exponents at the critical point $\Delta=2J$: We find
\begin{equation}
\IPR_0 \sim (V_2-V_{2\textrm{c}})^{\nu}
\qquad \textrm{and} \qquad 
\gamma \sim (V_2-V_{2\textrm{c}})^{\beta},
\end{equation}
with $\nu \simeq 0.33 \pm 0.015$ and $\beta \simeq 0.96 \pm 0.04$.

\begin{thebibliography}{10}
\bibitem{note:DOS}
{The quantity $n_\epsilon (E)/\epsilon$ may be interpreted as the density of
  states (DOS) for an energy resolution $\epsilon$. Because of the fractality
  of the energy spectrum, the DOS- $\lim_{\epsilon \rightarrow 0_+} n_\epsilon
  (E)/\epsilon$- is, however, ill defined (see below).}
\bibitem{note:DH}
{For a continuous (respectively\ discrete) spectrum, one finds $\DH=1$
  (respectively\ $0$). Intermediate values of $\DH$ are characteristic of a
  nontrivial self-similar behavior.}
\end{thebibliography}
\end{document}